\documentclass[superscriptaddress,twocolumn]{revtex4}

\usepackage{graphicx,color,url}
\definecolor{b}{rgb}{0,0,1.0}
\definecolor{r}{rgb}{0,0,0}
\definecolor{g}{rgb}{0,1,0}

\begin{document}

\newcommand{\SZFKI}{Research Institute for Solid State Physics and Optics,
 P.O. Box 49, H-1525 Budapest, Hungary}

\title{High speed imaging of traveling waves in a granular material during silo discharge}

\author{Tam\'as B\"orzs\"onyi}

\email{btamas@szfki.hu}
\affiliation{\SZFKI}
\author{Zsolt Kov\'acs}
\affiliation{\SZFKI}

\begin{abstract}
We report experimental observations of sound waves in a granular material
during resonant silo discharge called {\it silo music}. The grain motion was 
tracked by high speed imaging while the resonance of the silo was detected
by accelerometers and acoustic methods. The grains do not oscillate in phase 
at neighboring vertical locations, but information propagates upward in 
this system in the form of sound waves. We show that the wave velocity is not constant 
throughout the silo, but considerably increases towards the lower end of 
the system, suggesting increased pressure in this region, where 
the flow changes from cylindrical to converging flow.
In the upper part of the silo the wave velocity matches the sound velocity 
measured in the same material when standing (in the absence of flow).
Grain oscillations show a stick-slip character only in the upper part of the silo.
\end{abstract}

\maketitle

When a granular material is slowly discharged from a vertically aligned container 
(tube) one often observes vibrations accompanied by a booming low frequency 
(typically $50-200$ Hz) sound. This phenomenon can be observed in large silos 
and is often referred to as {\it silo music} or {\it silo honking}, 
but it can also easily be reproduced in smaller scale laboratory experiments 
\cite{te1999,buch2005,muqu2004,dhjo2006,wiru2008,nite2009,wite2010,tegu1993}. {\it Silo quaking}
can also be observed during discharge \cite{we2002,rowe2002} where short sharp bursts of motion are 
separated by longer periods of no movement. 

Although resonances during silo discharge are long known in industry  
and have recently been investigated thoroughly \cite{dhjo2006,wiru2008,nite2009,wite2010},
the phenomenon is not yet fully understood.
While one explanation is based on stick-slip motion of the grains near the silo walls 
\cite{muqu2004,buch2005,dhjo2006}, other authors argue that the phenomenon
originates from a dynamic interaction between the silo structure and the flowing material, 
where the source of the vibration is the transition zone between cylindrical and 
converging flow in the lower part of the system near the outlet 
\cite{wiru2008,nite2009,wite2010}. Recent laboratory 
experiments carefully investigated the frequency spectrum of the vibrations 
by piezoelectric accelerometers and microphones 
\cite{muqu2004,dhjo2006,wiru2008,nite2009,wite2010} and related these observations
of the calculated eigenfrequencies of the silo structure \cite{wiru2008} 
and the time evolution of the density of the material detected by a non-invasive method 
(Electrical Capacitance Tomography) \cite{nite2009,wite2010}. 
Optical imaging of 
{\it silo quaking} was reported in \cite{bupa2004} using a frame rate of 8.3 fps.
Here the predominant frequency of the bursts was about $1$ Hz,
which was significantly lower (by a factor of 2 or 4) than the frequency of the 
silo wall vibrations indicating no resonance in this case. Moreover, grain oscillations 
were in phase over the largest part of the silo wall, so grain motion could not be 
linked to wave propagation. 
This contradicts other measurements on silo quaking \cite{we2002,rowe2002} where 
the quake propagated upwards in the silo with a well defined velocity and growing amplitude, 
as detected by accelerometers embedded into the granular material. Very recent 
observations \cite{boca2010} on resonant pipe flow (the pipe is continously fed from above) 
detected upward propagating waves using piezoelectric sensors, and claim constant wave 
speed in the entire tube, meaning, that the
transition zone near the outlet has no special role in the formation of the resonance.
This contradicts other measurements \cite{wiru2008}, where manipulating the geometry 
of the transition zone allows for very effective reduction of the resonance.

In this Brief Report we present direct observations of the grain motion 
during resonant silo discharge by means of high speed imaging using a frame rate 
of 3000 fps. 
This non-invasive technique allows us to better characterize the spatial 
evolution of the system.
The aim of this study is to measure whether the grain oscillations show a 
stick-slip character near the walls, and if so, in which part of the silo? 
Are there regions in the silo, where the grains oscillate coherently in 
neighboring vertical positions, or not? What is the speed 
of wave propagation in the granular material during resonance? 

\begin{figure}[htb]
\includegraphics[width=\columnwidth]{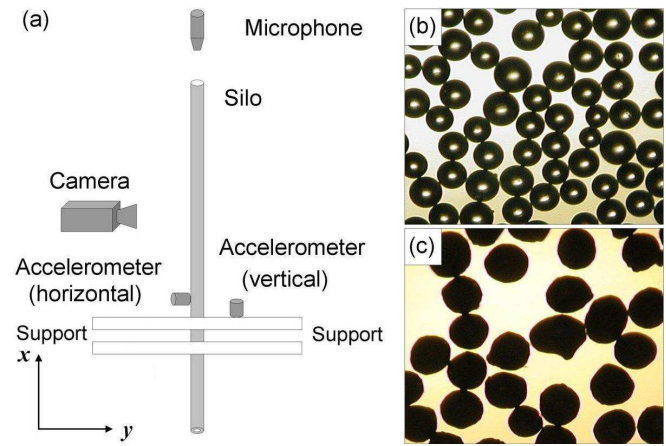}
\caption{(color online) (a) Schematic illustration of the experimental setup and   
(b)-(c) microscopic images of the materials used: (b) glass beads with mean diameter
of $d=0.15 \pm 0.05$ mm, (c) copper particles with $d=0.15 \pm 0.05$ mm.}
  \label{setup}
\end{figure}

In the present experiments the silo flow was realized in a vertically 
aligned glass tube of length $150$ cm (see Fig. \ref{setup}(a)). 
Two tubes were used with the inner diameters of $D=3.55$ cm and $2.3$ cm. 
We present results for two materials (i) nearly spherical copper particles
with average grain diameter of $d=0.15 \pm 0.05$ mm and (ii) spherical glass beads  
with $d=0.15 \pm 0.05$ mm (see Figs. \ref{setup}(b)-(c)). 
Qualitatively similar results were obtained with
quartz sand, granite and corundum of similar sizes, but in those cases the 
resonance was less pronounced.  
For the detection of the grain motion we used high speed cameras 
(Mikrotron CAMMC1310 and 1362) with a frame rate of 3000 fps.
The resonance was also detected by traditional methods: piezoelectric 
accelerometers mounted on the glass tube (for horizontal vibration), 
on the support of the tube (for vertical vibration), and a 
condenser microphone with large diaphragm (IMG Stageline ECM-140).
The silo discharge speed was varied by changing the diameter of the 
outlet $D_{out}$ at the bottom of the tube. As the main features of the
phenomenon we are focusing on do not depend on the precise value of 
$D_{out}$, for convenience a constant value $D_{out}=8$ mm was chosen for this study.

First we present the general features of the resonance obtained by 
piezoelectric accelerometers and the microphone. In one experiment the
above signals have been recorded during the initially full hopper discharges.
\begin{figure}[ht]
\includegraphics[width=\columnwidth]{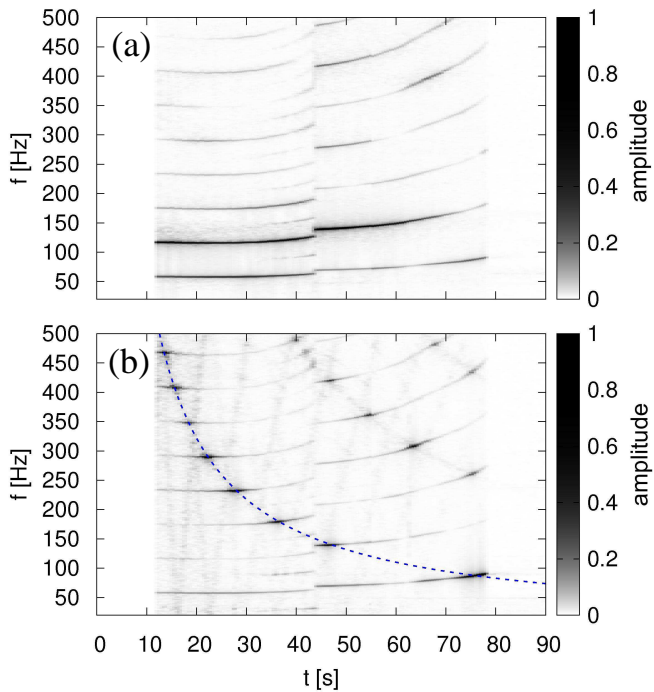}
\caption{(color online) Time evolution of the DFT power spectrum of the signals
obtained by the (a) accelerometer mounted on the tube, (b) microphone. Data obtained with copper 
particles  in the tube with the diameter of $D=3.55$ cm. 
The dashed line corresponds to the frequency of the standing 
wave (first harmonic) in the air column above the material.}
  \label{accel}
\end{figure}
This takes about 80 s and 120 s for the tube with $D=2.3$ cm and $D=3.55$ cm,
respectively. 
The power spectrum of these signals has been calculated using discrete Fourier
transform (DFT) with measurement windows of 0.5 s and is shown as a function
of time in Figs. \ref{accel}(a) and (b).
As it is seen in 
Fig. \ref{accel}(a) at the beginning of the process the system vibrates 
with a fundamental frequency of 
about 60 Hz, and this frequency slightly increases as the silo discharges.
At $t=44$ s there is a quick transition characterized by a jump in the
dominant frequency, but the trend of a slowly increasing frequency of the 
dominant mode persists until the resonance disappears at about $t=78$ s.
This picture is fully coherent with other observations 
\cite{wiru2008,nite2009,wite2010}
and can be explained as follows. The granular flow excites the eigenmodes of 
mechanical vibration of the tube itself and the frequency of these 
modes is slightly increasing in time as the material is flowing out 
of the system and the weight of the oscillating body decreases.
The time evolution of the system allows for changing the actually 
excited eigenmode, for which we have an example at $t=44$ s where 
we see a jump in the dominant resonant frequency.

The signal picked up by the microphone shows the same characteristics
as that of the accelerometer, but in addition to the eigenmode with 
slightly increasing frequency we find a resonance with decreasing 
frequency. This is the sign of standing sound waves in the air column
in the tube above the granular material. As the length $l$ of this column 
is increasing with time (as the silo discharges) the frequency of the 
dominant mode is decreasing. The amplitude of this signal gets strong
when it resonates with the "driving" frequency. The dashed line 
corresponds to the frequency of the standing wave (first harmonic) according to 
the formula of 
$f(t)=\frac{c}{\lambda(t)} = \frac{c}{4l(t)}= \frac{1}{4} \cdot\frac{c}{v_s t+0.3D}$
where $v_s$ is the velocity by which the top surface of the granular bed 
is sinking, $c$ is the velocity of sound in air
and $0.3 D$ is the length correction for the open end of the tube \cite{ra1945}.
The value of $v_s$ has been 
measured in an additional experiment, where the whole process was 
recorded optically and the position of the top surface was determined
by digital image analysis. Several measurements confirmed, that $v_s$
is nearly independent of time except near the very end of the process where $v_s$ 
slightly increased.
 
\begin{figure}[ht]
\includegraphics[width=\columnwidth]{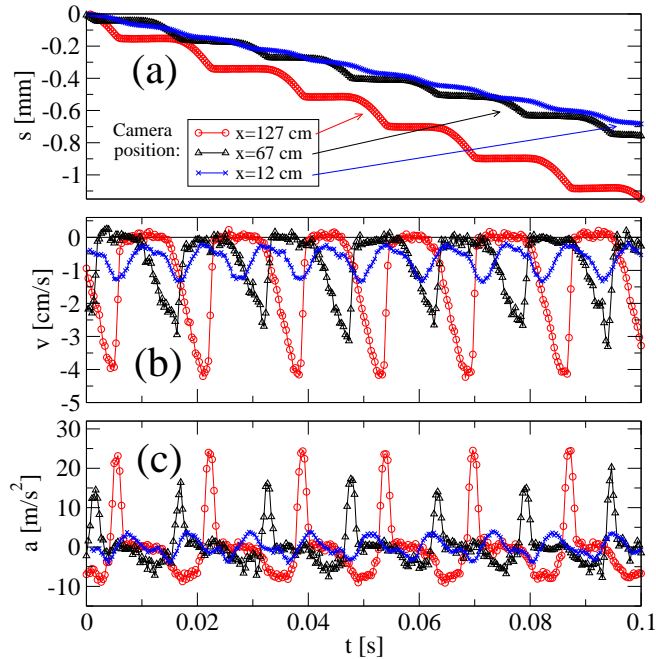}
\caption{(color online) (a) Grain position, (b) velocity 
and (c) acceleration as a function of time taken at three heights 
$x=12$ cm ($\times$), $x=67$ ($\triangle$) cm and $x=127$ ($\circ$) cm measured 
from the bottom of the tube. Data taken with copper particles in the tube with 
the diameter of $D=3.55$ cm.
}
  \label{motion}
\end{figure}

In the following we will focus on the data obtained with high speed imaging.
First we show how the grains move at a selected height $x$. For this we took 
images of a vertical stripe of $1 \times 0.1$ cm and determined the displacement 
of the particles by correlating subsequent images. This procedure was repeated at 
various heights. In Fig. \ref{motion} three
examples are shown, where Fig. \ref{motion}(a) shows the position $s$ of the particles
as a function of time, while (b) and (c) show the corresponding velocity $v$ and 
acceleration $a$. The three measurements correspond to $x=12$ cm, $x=67$ cm and $x=127$ cm
measured from the bottom of the tube. As it is seen, the grains
move downwards with oscillating velocity. At the top and middle of the tube 
during each period they come to a full stop, remain standing for a considerable portion 
of the period and then 
accelerate downwards. Their acceleration almost reaches  
the value corresponding to free fall ($g=-9.81$ m/s$^2$) near the top of the tube, but is 
lower in the middle of the tube. At the bottom of the tube one can still see
oscillations in the grain velocity but the grains do not come to a full stop and their 
acceleration does not come close to $g$. 
The average grain velocity near the walls $v_{av}$ (i.e. the slope of the curves in 
Fig. \ref{motion}(a)) is smaller than the velocity $v_s$ by which the top surface of 
the granular bed is sinking. The corresponding three values are $v_{av}/v_s = 0.54$,
0.59 and 0.89 for $x=12$ cm, $67$ cm and $127$ cm, respectively. This means that no 
perfect plug flow develops, i.e. the grains move slower near the walls than in the center of 
the tube even at the uppermost part of the material. 

\begin{figure}[ht]
\includegraphics[width=\columnwidth]{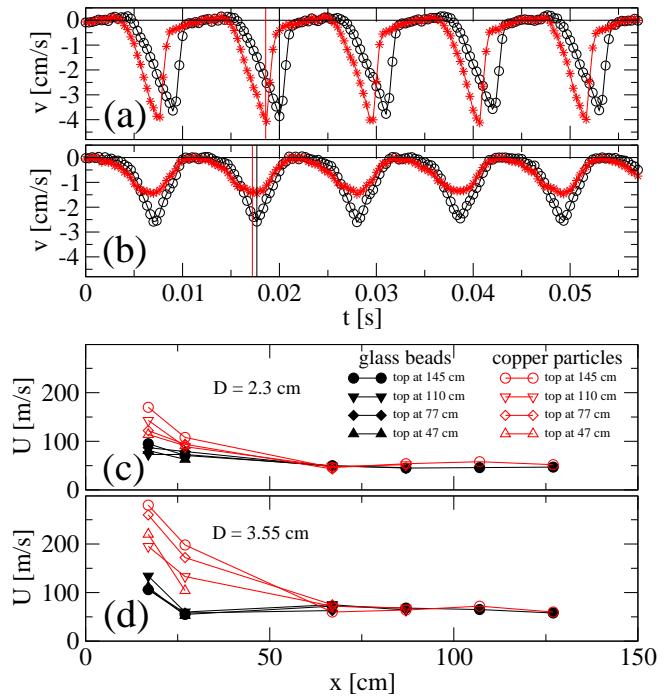}
\caption{(color online) Grain velocity $v$ as a function of time at two nearby 
locations at (a) $x=102$ cm ($\star$) and $x=113$ cm ($\circ$) and (b) $x=22$ cm ($\star$) 
and $x=33$ cm ($\circ$). The time lag between the curves is a measure of the corresponding 
sound velocity.   
(c)-(d) Velocity of sound waves as a function of location $x$ in the tube 
with (c) $D=2.3$ cm and (d) $D=3.55$ cm. The curves correspond to different 
stages of the process, when the top of the material was at $x=145$ cm ($\circ$), 
$x=110$ cm ($\nabla$), $x=77$ cm ($\diamond$) and $x=47$ cm ($\triangle$). 
Data for copper with open symbols, data for glass beads with solid symbols.
}
  \label{waves}
\end{figure}

In order to measure whether or not the grains at different heights oscillate in phase with each 
other, a longer vertical stripe of $13.2 \times 0.5$ cm was monitored. Two $2.2$ cm long
segments were analyzed at the upper and lower parts of this image.  As it is seen in 
Figs. \ref{waves}(a)-(b) there is a phase shift between these signals (indicated by vertical lines)
implying that the grains
in neighboring segments do not oscillate in phase with each other, but waves
are propagating upwards in the system. The phase shift between the neighboring peaks
in Fig. \ref{waves}(a) corresponds to a wave velocity of $U=54$ m/s, while in the lower part 
of the tube (Fig. \ref{waves}(b)) a larger wave velocity (smaller phase shift) is observed. 
The velocity $U$ of the upward propagating waves has been measured at 6 different
vertical positions. The resulting $U(x)$ curves are shown in figs. \ref{waves}(c) and (d)
for the two tubes with $D=2.3$ cm and $3.55$ cm, respectively. This measurement has 
been repeated at four different stages of the process, where the top of the material was
at $x=145$ cm, $x=110$ cm, $x=77$ cm and $x=47$ cm.
The wave velocity $U$ appears to be constant in the upper part of the tube. 
In the smaller tube ($D=2.3$ cm) we get $U=44 \pm 6$ m/s for glass beads and $U=54\pm6$ m/s for 
the copper particles, while in the larger tube ($D=3.55$ cm) $U=65\pm6$ m/s for both materials.
These values are similar to those previously observed for the velocity of
upward propagating pulses detected during {\it "silo quaking"} \cite{we2002}.
There single rarefaction
waves (quakes) propagated upwards in the system with a velocity of $74\pm15$ m/s and
$135\pm20$ m/s in the two Perspex tubes with the diameters of $9$ cm and $30$ cm respectively.
The fact that we find larger value of $U$ in the larger tube also agrees with the above data 
obtained for quakes and is in accordance with very recent observations on continous pipe flow, 
where the velocity of sound waves was demonstrated to increase with increasing tube diameter 
\cite{boca2010}. Our more detailed spatial characterization however revealed
(Figs. \ref{waves}(c)-(d)), that the wave velocity 
is not constant throughout the tube but it increases towards the bottom of the tube. 
The larger value of $U$ measured at the lower part of the tube means a smaller phase lag 
between the oscillation of neighboring particles, which, in accordance with other results 
\cite{wiru2008,wite2010}, shows that the resonance originates from that part of 
the tube where the flow changes from cylindrical to converging flow (at $x\approx4D$)
leading to strong density oscillations.

An additional experiment was done to measure the velocity of sound waves in 
the standing material (without flow) using the smaller tube ($D=2.3$ cm). In this 
measurement we placed 2 piezoelectric accelerometers in the granular material at a 
distance $dx$ above each other, 
and perturbed the granular bed at the lower end of the tube with a pneumatically generated
single pulse. This was done using a smaller (flexible) tube with a rubber membrane at 
the end, which was placed in the lower end of the silo. 
\begin{figure}[ht]
\includegraphics[width=7cm]{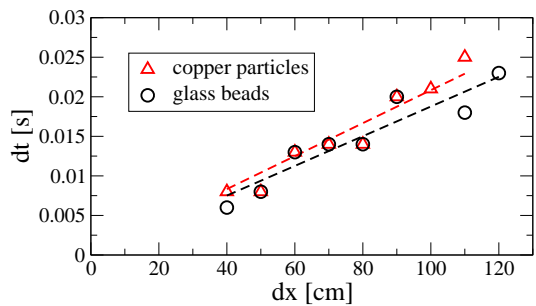}
\caption{(color online) The elapsed time $dt$ during a single pulse traveled between two 
piezoelectric accelerometers placed in the static material at a distance $dx$ 
for the case of copper ($\triangle$) and glass beads ($\circ$) in the tube with $D=2.33$ cm.
The slope of the linear fits (shown with dashed lines) gives $U_s$ the velocity of sound waves
in the standing material.
}
\label{standing}
\end{figure}
We measured the time $dt$ needed for the pulse to travel between the two accelerometers as a 
function of $dx$.
The slope of the resulting curves 
(presented in Fig. \ref{standing}) gives for the 
speed of sound waves in the standing material $U_s = 53\pm5$ m/s 
for glass beads and $U_s=48\pm5$ m/s for the copper particles. 
These values agree nicely with the ones discussed above observed in the same tube during 
flow (resonance) and are similar to other experimentally obtained values for the 
sound speed in glass beads ($60-240$ m/s \cite{lina1993}; $40-60$ m/s \cite{boan2008})
or sand ($40$ m/s \cite{an2004}), or even using 
much softer photoelastic particles ($100$ m/s \cite{owda2010}).  
These sound speed measurements were done in the absence of strong external load and also 
revealed, that the result can significantly depend on the measurement method \cite{lina1993}. 
More importantly, recent investigations uncovered, that in samples under external uniaxial load, the
velocity of compressional waves is sensitive to stress induced anisotropy, leading to larger $U$
perpendicular to the applied load than parallel to it \cite{khji2010}.
Also, $U$ increases more gradually with pressure $P$ ($U\propto P^{1/4}$) at lower pressures, 
compared to the generally observed $U\propto P^{1/6}$ dependence at higher $P$ \cite{jica1999}.
Both of these observations \cite{jica1999,khji2010} indicate, that the larger wave velocity 
measured during resonant flow in the lower part of the tube suggests increased radial pressure
in this region, where the velocity gradient changes due to the proximity of the outlet. 

In summary, we studied a resonance phenomenon called {\it "silo music"}
during silo discharge. 
We have visualized the oscillations in grain motion with high speed imaging, 
and detected that grain oscillations are not coherent when compared at 
different heights, but there is a phase lag. Information propagated upward in the system in 
the form of sound waves. One of our major findings is, that the wave velocity is
not constant throughout the silo, but considerably increases towards the lower end of the system (corresponding to 
a more coherent motion of the grains). This underlines the importance of the transition 
zone in the lower end of the silo,  
where the flow changes from cylindrical to converging flow 
and is in accordance with those observations 
\cite{wite2010,wiru2008} which suggest that the source of the resonance phenomenon 
is the generation of strong stress oscillations in this region.
As shown, the amplitude of the grain oscillations increases with height and leads to
stick-slip motion in the upper part of the silo. This supports
the argument that appropriate friction with the walls is crucial for the amplification of the waves
\cite{we2003,we2002,we2002b,boca2010} and could be a necessity for strong resonance as argued by other 
authors \cite{muqu2004,dhjo2006,buch2005}, which is also supported by the fact that 
increasing the roughness of the silo walls (i.e. suppressing stick-slip) 
effectively reduces the resonance \cite{tegu1993,te1999}. 
Finally, we have also shown that the velocity of the waves in the upper part of the silo 
matches the sound velocity measured in the same material when standing (in the absence of flow).

The authors are thankful for discussions with Kesava Rao, who 
motivated us to study this problem.
We thank A. Konya for critical reading of this manuscript.


\begin{thebibliography}{99}

\bibitem{wite2010} K. Wilde, J. Tejchman, M. Rucka, M. Niedostatkiewicz, 
Powder Technology {\bf 198}, 38 (2010).

\bibitem{nite2009} M. Niedostatkiewicz, J. Tejchman, Z. Chaniecki, K. Grudzie\'n, 
Chemical Engineering Science {\bf 64}, 20 (2009).

\bibitem{wiru2008} K. Wilde, M. Rucka, J. Tejchman, Powder Technology {\bf 186}, 113 (2008).

\bibitem{dhjo2006} M.L. Dhoriyani, K.K. Jonnalagadda, R.K. Kandikatla, K.K. Rao, 
Powder Technology {\bf 167}, 55 (2006).

\bibitem{muqu2004} B.K. Muite, F.S. Quinn, S. Sundaresan, K.K. Rao, 
Powder Technology {\bf 145}, 190 (2004).

\bibitem{buch2005} J.M. Buick, J. Chavez-Sagarnaga, Z. Zhing, J.Y. Ooi, D.M. Pankaj, D.M. Cambell,
C.A. Greated, Journal of Engineering Mechanics, ASCE {\bf 131}, 299 (2005).

\bibitem{te1999} J. Tejchman, Powder Technology {\bf 106}, 7 (1999).

\bibitem{tegu1993} J. Tejchman and G. Gudehus, Powder Technology {\bf 76}, 201 (1993).

\bibitem{we2002} C. Wensrich, Powder Technology {\bf 127}, 87 (2002). 

\bibitem{rowe2002} A.W. Roberts and C.M. Wensrich, Chem. Eng. Sci. {\bf 57}, 295 (2002).

\bibitem{bupa2004} J.M. Buick, Y. Pankai, J.Y. Ooi, J. Chavez-Sagarnaga, A. Pearce, 
G. Houghton, Journal of Physics. D: Applied Physics {\bf 37}, 2751 (2004).

\bibitem{boca2010} L. Bonneau, T. Catelin-Jullien and B. Andreotti, 
Phys. Rev. E {\bf 82}, 011309 (2010).

\bibitem{ra1945} J.W.S. Rayleigh, The Theory of Sound, Dover Publications (1945).

\bibitem{lina1993}  C.H. Liu and S.R. Nagel, Phys. Rev. B {\bf 48}, 15646 (1993); 
C.H. Liu, Phys. Rev. B {\bf 50}, 782 (1994). 

\bibitem{an2004}  B. Andreotti, Phys. Rev. Lett. {\bf 93}, 238001 (2004). 

\bibitem{boan2008}  L. Bonneau, B. Andreotti and E. Clement, Phys. Rev. Lett. 
{\bf 101}, 118001 (2008). 

\bibitem{owda2010}  E.T. Owens and K.E. Daniels, unpublished. 

\bibitem{jica1999}  X. Jia, C. Caroli and B. Velicky, Phys. Rev. Lett. 
{\bf 82}, 1863 (1999).

\bibitem{khji2010} Y. Khidas and X. Jia, Phys. Rev. E {\bf 81}, 021303 (2010).

\bibitem{we2003} C. Wensrich, Int. J. Mech. Sci. {\bf 45}, 541 (2003).

\bibitem{we2002b} C. Wensrich, Powder Technology {\bf 126}, 1 (2002).

\end{thebibliography}
\end{document}